\documentclass[preprint,12pt,authoryear]{elsarticle} 

\usepackage[top=2cm,bottom=2.5cm,left=2.0cm,right=2.0cm]{geometry}

\usepackage{setspace}

\usepackage{placeins} 

\usepackage{amssymb}
\usepackage{amsmath}

\usepackage{multirow} 
\usepackage{makecell} 

\usepackage{lineno}

\usepackage{xcolor} 

\usepackage{hyperref}
\hypersetup{
hidelinks = true, 
    }  

\journal{Journal of the Mechanics and Physics of Solids}

\begin{document}

\begin{frontmatter}



\title{\textbf{~\\ ~\\Origami of multi-layered spaced sheets}} 


\author[1]{Guowei Wayne Tu}
\ead{guoweitu@umich.edu}
\affiliation[1]{organization={Deployable and Reconfigurable Structures Laboratory, Department of Civil and Environmental Engineering},
            addressline={University of Michigan}, 
            city={Ann Arbor},
            postcode={48109}, 
            state={MI},
            country={USA}}

\author[1,2]{Evgueni T. Filipov\corref{cor1}}
\ead{filipov@umich.edu}
\affiliation[2]{organization={Deployable and Reconfigurable Structures Laboratory, Department of Mechanical Engineering},
            addressline={University of Michigan}, 
            city={Ann Arbor},
            postcode={48109}, 
            state={MI},
            country={USA}}
\cortext[cor1]{Corresponding author}

\begin{abstract}
Two-dimensional (2D) origami tessellations such as the Miura-ori are often generalized to build three-dimensional (3D) architected materials with sandwich or cellular structures. However, such 3D blocks are densely packed with continuity of the internal material, while for many engineering structures with multi-physical functionality, it is necessary to have thin sheets that are separately spaced and sparsely connected. This work presents a framework for the design and analysis of multi-layered spaced origami, which provides an origami solution for 3D structures where multiple flat sheets are intentionally spaced apart. We connect Miura-ori sheets with sparsely installed thin-sheet parallelogram-like linkages. To explore how this connectivity approach affects the behavior of the origami system, we model the rigid-folding kinematics using analytic trigonometry and rigid-body transformations, and we characterize the elastic-folding mechanics by generalizing a reduced order bar and hinge model for these 3D assemblies. The orientation of the linkages in the multi-layered spaced origami determines which of three folding paths the system will follow including a flat foldable type, a self-locking type, and a double-branch type. When the origami is flat foldable, a maximized packing ratio and a uniform in-plane shear stiffness can be achieved by strategically choosing the link orientation. We show possible applications by demonstrating how the multi-layered spaced origami can be used to build deployable acoustic cloaks and heat shields. 
\end{abstract}




\end{frontmatter}

\begin{spacing}{1.2} 


\section{Introduction}\label{sec:Intro}
A large number of origami tessellations, such as the Miura-ori pattern, the Yoshimura pattern, and the Kresling pattern, can be used to build reconfigurable thin sheet structures \citep{evans2015lattice,callens2018flat,dudte2016programming}. The Miura-ori pattern \citep{miura2009science} is an especially popular pattern that has been widely explored and generalized because it is developable, and enables flat and rigid foldability of large surfaces. \cite{sareh2015design,sareh2015design_2} introduced methodologies for both isomorphic and non-isomorphic Miura-ori descendant design; \cite{eidini2015unraveling} proposed the zigzag folded sheets to expand on the design space of Miura-ori while preserving the remarkable properties of it; \cite{kamrava2018programmable} designed the Miura-ori-based origami strings, which are slender sheet-like structures with programmable deployment trajectories. Two-dimensional (2D) origami sheets have also been generalized to build three-dimensional (3D) densely-arranged cellular blocks: the stacked Miura sandwich plates \citep{schenk2013geometry,chen2023computational,schenk2014novel}, assembled zipper tubes \citep{filipov2015origami,li2016recoverable,webb2024asymmetric}, self-locking honeycombs \citep{gao2022origami,fang2018programmable,fang2016self}, and architected multi-mode cells \citep{liu2023digitized,jamalimehr2022rigidly,fang2017dynamics} are some good examples of 3D cellular origami based on Miura-ori patterns. 

These Miura-ori-inspired 2D origami sheets and 3D origami blocks can offer unprecedented mechanical properties, including ultra-high stiffness \citep{zhu2023soft,miyazawa2021heterogeneous}, tunable Poisson's ratio \citep{misseroni2022experimental,yasuda2015reentrant,pratapa2019geometric}, multi-stability \citep{liu2022triclinic,tao2022asymmetric}, self-locking \citep{fang2018programmable,ye2023multimaterial}, and fast deployability \citep{baek2020ladybird,zhao2023deployable}. Engineers have used these properties of the Miura-ori to build materials and structures across disciplines and scales, ranging from meter-scale deployable shelters \citep{thrall2014accordion,norman2017origami} to centimeter-scale tunable acoustic wave guides \citep{hathcock2021origami,bentley2022acoustic} and micrometer-scale electrothermal robots \citep{zhu2020elastically,wu2020self}.

\begin{figure}[!htb] 
\centering
\makebox[0pt]{\includegraphics[scale=1]{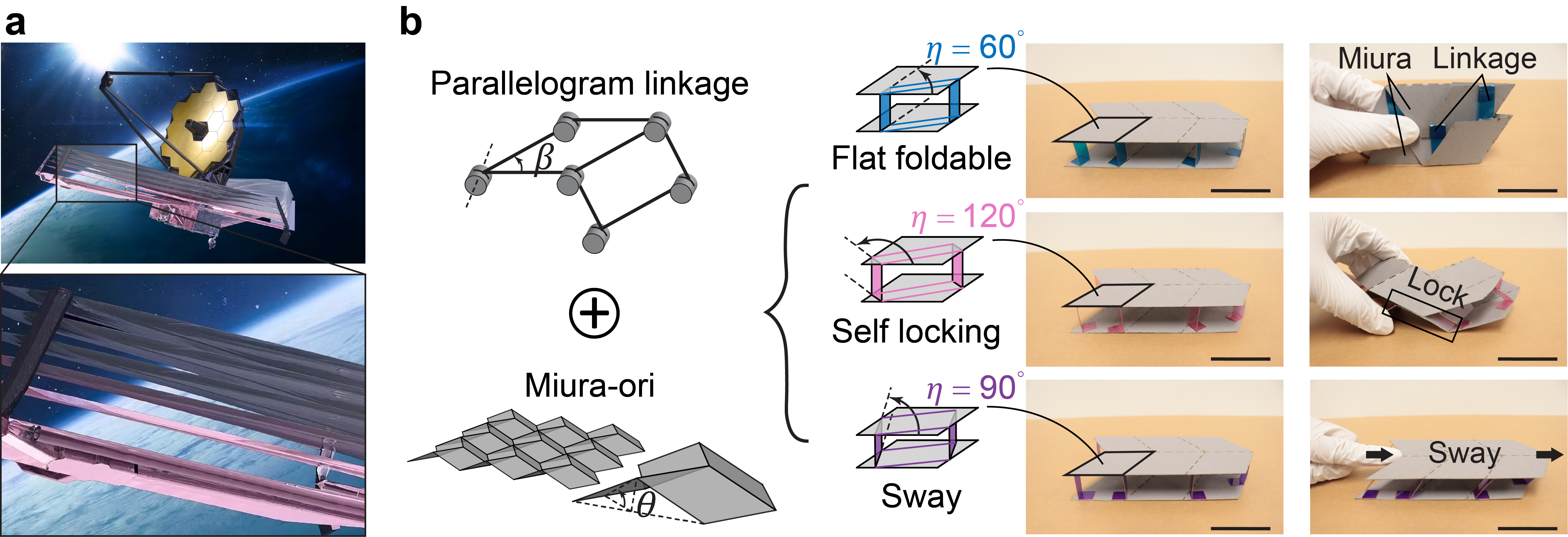}} 
\caption{Overview of multi-layered spaced origami explored in this work. (a) The heat shield of the James Webb Space Telescope is a typical structure where multiple sheets are intentionally spaced; Image source: \cite{JAMESWEBB}; (b) The multi-layered sheets are created by connecting Miura-ori with thin-sheet parallelogram linkages. The link orientation angle $\eta$ affects the kinematic and mechanical behavior of the systems (scale bars are 4 cm).}\label{fig:Intro}
\end{figure}

Despite the outstanding properties and broad range of applications of the Miura-ori, existing designs cannot achieve multi-layered and sparsely-connected sheets which are necessary for creating engineering systems such as heat shields, parallel plate capacitors, and acoustic cloaks (see Fig.~\ref{fig:Intro}(a)). To function properly, these structures require a considerable amount of space between parallel 2D sheets, whereas existing 3D origami blocks such as stacked Miura sandwiches \citep{schenk2013geometry} are densely packed with continuous connectivity between the spaced sheets. The lack of an origami solution for multi-layered spaced sheets limits the applications of adaptive functional structures for multi-physical purposes in engineering. 

To address this issue, we develop a general framework for the design and analysis of \textit{multi-layered spaced origami}, where multiple flat sheets are intentionally spaced apart and are only connected by sparsely installed links. In Sec.~\ref{sec:Design}, we demonstrated how to design and fabricate multi-layered spaced origami by connecting individual Miura-ori sheets with thin-sheet parallelogram linkages (see Fig.~\ref{fig:Intro}(b) for an overview). Sec.~\ref{sec:Modeling} presents an analytical model based on rigid body transformations to capture the folding kinematics, and adapts a bar and hinge model to simulate elastic-folding behaviors. Then in Sec.~\ref{sec:Properties}, we studied the kinematic and mechanical properties of the multi-layered spaced origami using the developed simulation tools. We explored how specific design parameters lead to distinct folding paths, different packing ratios, and isotropic or anisotropic shear stiffness. In Sec.~\ref{sec:Applications}, we designed and simulated meter-scale acoustic cloaks and adjustable heat shields using our multi-layered spaced origami to show potential applications. Sec.~\ref{sec:Conclusions} provides conclusions.


\section{Design and fabrication of multi-layered spaced origami}\label{sec:Design}
\subsection{Origami design and geometric definitions}\label{sec:GeometryDef}
Figure~\ref{fig:Geometry}(a) shows the design of the \textit{basic assembly} for multi-layered spaced origami. We use rectangular links to connect two Miura-ori sheets. Multiple strips with two folds are attached to the Miura-ori sheets so that each strip acts as a link and each crease acts as a hinge. All the links are placed with the same \textit{link orientation angle} $\eta$, and are symmetric about the s-s axis, the axis of symmetry of the Miura-ori (see Fig.~\ref{fig:Geometry}(c)). The entire structure can be seen as a double-layered Miura-ori coupled with parallelogram mechanisms. 

Here, we define the coordinate systems and all the geometric parameters of our origami. As shown in Fig.~\ref{fig:Geometry}(b, c), the global origin $o$ is placed at the upper left corner of the bottom Miura sheet, and the global Cartesian system ($x, y, z$) aligns with three orthogonal axes of the Miura-ori. The geometry of the Miura unit cell, shown in Fig.~\ref{fig:Geometry}(b, c), is defined by the two lengths \textit{a} and \textit{b}, and the \textit{Miura sector angle} $\gamma$. The folded configuration of the Miura unit cell is then defined by the \textit{folding angle} $\theta$, which is the dihedral angle between the folded Miura panel and the $x$-$y$ plane \citep{schenk2013geometry}. With all the parameters defined, we can locate nine vertices of a Miura unit and thus determine the shape of the two Miura-ori sheets at any folded state. 

Each link is a rectangle with width $w$ and length $d$. For our basic assembly, eight links are installed symmetrically about the s-s axis with the same orientation angle $\eta$, as mentioned earlier. To locate the links during the folding of the basic assembly, we establish local coordinate systems ($x_{i}, y_{i}, z_{i}$) to describe the rigid body motion of the links (where $i$ is a number that labels the link; see Fig.~\ref{fig:Geometry}(d)). The local origin $o_{i}$ is placed at the outer vertex of the link, and the $x_{i}$ axis (the axis of rotation of the link) is defined to be along the hinge and pointing toward the inside of the Miura panel. The $y_{i}$ axis is the other orthogonal axis lying on the Miura panel, while the $z_{i}$ axis is an orthogonal axis that points out from the panel. Within the local coordinate system, we define the \textit{link rotation angle} $\beta _{i}$ which is the dihedral angle between the Miura panel and the link panel. With the global coordinates of the local origin $o_{i}$ and the value of the link rotation angle $\beta _{i}$, we can determine the geometry of the links in 3D space. Later, we will prove that all $\beta _{i}$ have the same value. We summarize all the design and kinematic control parameters of the basic assembly in Table~\ref{tab:ParaAssembly}. 

Although the placement of the links in our basic assembly follows a specific pattern, these links can be placed arbitrarily between the two origami sheets. That is, an arbitrary number of links can be installed at arbitrary positions as long as they remain consistent in their orientation angle $\eta$ (see Fig.~\ref{fig:Geometry}(f) for an example). While all links must have the same length $d$, they can have an arbitrary width $w$ and thickness $t_2$. These rules enhance the flexibility of the multi-layered origami design, which is especially useful when the performance of origami devices is sensitive to the position and density of the inner fillings. We used paper sheets and plastic strips to show the compatible folding of these systems, but our choice of materials does not affect the kinematics of the structure. We will revisit these points in later sections. 

\begin{table}[htbp]
  \centering
\caption{Design and kinematic control parameters of the basic assembly.}
\label{tab:ParaAssembly}
  \begin{tabular}{crl}
  \hline
Category&Type& \multicolumn{1}{l}{Items} \\
   \hline 
\multirow{3}{*}{Design parameters}  & Miura  & $a,b,\gamma,t_1$ -- Miura panel side lengths, sector angle, and thickness\\
                    & Link  & $d,w,t_2$ -- Link length (spacing), width, and thickness \\
                    & Connection  & $\eta$ -- Link orientation angle \\                
  \hline
\multirow{2}{*}{\makecell{Kinematic control\\parameters}}  & Miura  & $\theta$ -- Miura folding angle \\
                    & Link  & $\beta$ -- Link rotation angle\\
  \hline                 
\end{tabular}
\end{table}

\subsection{Fabrication}
\begin{figure}[!htb] 
\centering
\makebox[0pt]{\includegraphics[scale=1]{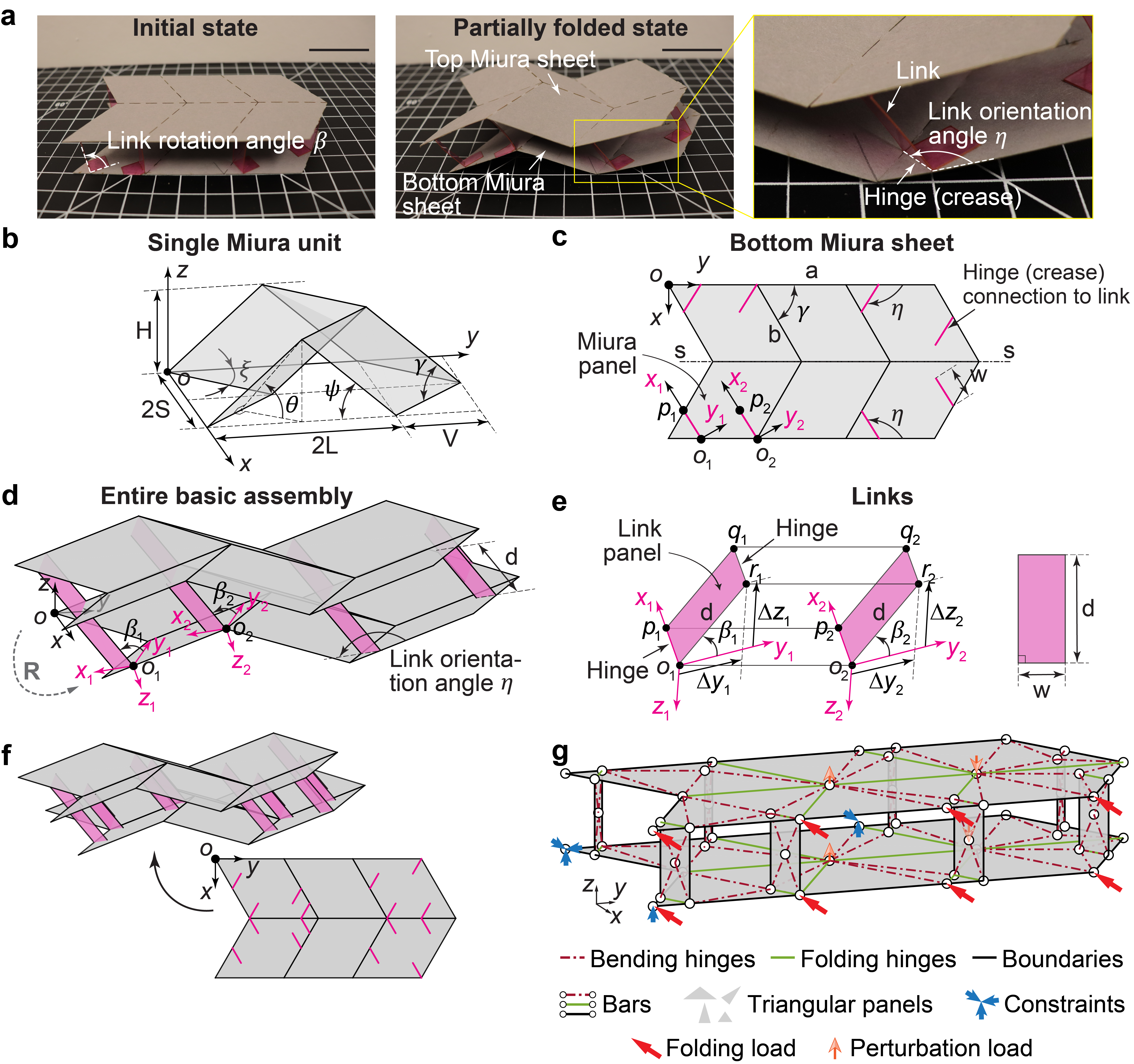}} 
\caption{Geometry, coordinates, and bar \& hinge mesh of the basic multi-layered assembly. (a) The initial deployed state, a partially folded state, and a close-up of a linkage connector (scale bars are 4 cm); (b) Geometry of a single Miura unit; (c) Geometry of the bottom Miura sheet with markings for the positions and orientations of the eight links; (d) Geometry of the entire basic assembly; (e) Geometry of two of the eight links in (d); (f) Another pattern for link installation; links can be installed at arbitrary locations on alternating panels along $y$ axis, as long as links have the same orientation angle $\eta$; (g) Bar \& hinge mesh for the basic assembly with boundary conditions and loads to achieve folding (the perturbation load is a follower load constantly in parallel with the incremental displacement of the corresponding nodes during the search for equilibrium paths).}\label{fig:Geometry}
\end{figure}

The fabrication of the basic assembly involves three main steps. We first perforate two sheets with a Universal Laser System (VLS 6) to form the crease pattern for the Miura-ori. The sheets are made from 0.8 mm ($t_1$)-thick paperboard and each sheet consists of one and a half Miura units (see Fig.~\ref{fig:Geometry}(b, c)). We then use the same system to make 0.6 mm ($t_2$)-thick laser-cut Mylar\textsuperscript{®} strips with two creases. The middle segment of the strip acts as a link, and the two ends are connection tabs. Finally, we use Scotch\textsuperscript{®} clear glue to bond the sheets and strips via the tabs at the pre-defined positions, which are marked on Miura sheets by laser engraving when everything is at the flat state. We manually fold the entire assembly into a 3D shape after the glue cures. 

By varying the orientation of the links and other design parameters, we have countless ways to connect the sheets and the linkages. How are those combinations different and can we classify them? To answer this question, we model our basic assembly in the next section. 


\section{Modeling and testing approach}\label{sec:Modeling}
In this section, we detail our strategy for modeling and experimental testing of the designed multi-layered spaced origami. We focus on one single basic assembly, yet the same strategy applies to extended cellular structures. 
\subsection{Modeling of the rigid-folding kinematics}\label{sec:Kinematics}
First, we treat the folding of the basic assemblies as a rigid body motion problem and model the rigid-folding kinematics of the system. Based on the geometry as defined in Fig.~\ref{fig:Geometry}, a question can be raised: What conditions have to be met for our basic assembly to be rigid-foldable? The rigid folding of the Miura-ori sheets and links is already guaranteed, so all we need is the kinematic compatibility between the sheets and links. The configuration of the Miura-ori (which can be treated as a single-degree-of-freedom (S-DOF) mechanism) is controlled by the folding angle $\theta$, while the configuration of the links is controlled by the link rotation angle $\beta _{i}$. Thus, $\theta$ and $\beta _{i}$ have to be coupled because the local coordinate system ($x_i, y_i, z_i$) translates and rotates in the global coordinate system ($x, y, z$) as the bottom Miura sheet is folded. Therefore, the compatibility condition boils down to a quantitative relationship between $\theta$ and $\beta _{i}$. 

To determine this relationship, we first give two premises about the rigid body motion of the origami: 1) the length of any edge remains constant during folding; 2) any geometric relationship within an origami panel (Miura or link panel)\footnote{An origami panel is a polygon connected to adjacent polygons by pre-defined creases.} remains unchanged during folding. We name the four vertices of `link $i$' as $o_i, p_i, q_i$ and $r_i$, and we focus on the two links connected to the same Miura panel (see Fig.~\ref{fig:Geometry}(d, e)). Based on the two premises, the following mathematical derivation can be given: 
\begin{equation}\label{eqn:Geometry_1}
\left. \begin{array}{r}
\overline {{o_1}{p_1}} \parallel \overline {{r_1}{q_1}} \ \text{and} \ {\rm{ }}\overline {{o_2}{p_2}} \parallel \overline {{r_2}{q_2}} {\text{ (by Premise \#2)}}\\
\overline {{o_1}{p_1}} \parallel \overline {{o_2}{p_2}} {\text{ (by Premise \#2)}}
\end{array} \right\} \Rightarrow \overline {{o_1}{p_1}} \parallel \overline {{r_1}{q_1}} \parallel \overline {{o_2}{p_2}} \parallel \overline {{r_2}{q_2}} ,
\end{equation}
\begin{equation}\label{eqn:Geometry_2}
\begin{array}{c}
\left. \begin{array}{r}
\overrightarrow {{o_1}{x_1}} \parallel \overrightarrow {{o_2}{x_2}} \ \text{and} \ {\rm{ }}\overrightarrow {{o_1}{y_1}} \parallel \overrightarrow {{o_2}{y_2}} {\text{ (by Premise \#2)}}\\
\left| {{o_1}{o_2}} \right| = \left| {{r_1}{r_2}} \right| = {\text{constant (by Premise \#1)}}\\
\left| {{o_1}{r_1}} \right| = \left| {{o_2}{r_2}} \right| = {\text{d (by Premise \#1)}}
\end{array} \right\} \Rightarrow {\angle \beta _1} = {\angle \beta _2},\\
{\angle \beta _1} = {\angle \beta _2} \Rightarrow \sqsubset\!\sqsupset {o_1}{p_1}{q_1}{r_1}\parallel \sqsubset\!\sqsupset {o_2}{p_2}{q_2}{r_2} \Rightarrow \overline {{o_1}{r_1}} \parallel \overline {{o_2}{r_2}} ,\\
\left. \begin{array}{r}
\overline {{o_1}{r_1}} \parallel \overline {{o_2}{r_2}} \\
\left| {{o_1}{r_1}} \right| = \left| {{o_2}{r_2}} \right| = {\rm{d}}
\end{array} \right\} \Rightarrow \sqsubset\!\sqsupset {o_1}{o_2}{r_2}{r_1} {\text{ is a parallelogram}} \Rightarrow \overline {{o_1}{o_2}} \parallel \overline {{r_1}{r_2}} ,\\
\left. \begin{array}{r}
\overline {{o_1}{o_2}} \parallel \overline {{r_1}{r_2}} \\
\overline {{o_1}{o_2}} \parallel \overline {{p_1}{p_2}} \ \text{and} \ {\rm{ }}\overline {{r_1}{r_2}} \parallel \overline {{q_1}{q_2}} {\text{ (by Premise \#2)}}
\end{array} \right\} \Rightarrow \overline {{o_1}{o_2}} \parallel \overline {{r_1}{r_2}} \parallel \overline {{p_1}{p_2}} \parallel \overline {{q_1}{q_2}},
\end{array}
\end{equation}
where $\overline {AB}$ represents a line determined by points $A$ and $B$, $\overrightarrow {AB}$ represents a vector pointing from point $A$ to point $B$, $\left| {AB} \right|$ represents the length of a line segment connecting points $A$ and $B$, $\angle \alpha$ represents the value of angle $\alpha$, and $\sqsubset\!\sqsupset ABCD$ represents a quadrilateral connecting points $A,B,C$ and $D$ in order. Combining the outcomes from Eqs.~\ref{eqn:Geometry_1} and \ref{eqn:Geometry_2}, we have
\begin{equation}\label{eqn:Geometry_3}
\left. \begin{array}{r}
\overline {{o_1}{p_1}} \parallel \overline {{r_1}{q_1}} \parallel \overline {{o_2}{p_2}} \parallel \overline {{r_2}{q_2}} \\
\overline {{o_1}{o_2}} \parallel \overline {{r_1}{r_2}} \parallel \overline {{p_1}{p_2}} \parallel \overline {{q_1}{q_2}} 
\end{array} \right\} \Rightarrow \sqsubset\!\sqsupset {o_1}{o_2}{p_2}{p_1}\parallel \sqsubset\!\sqsupset {r_1}{r_2}{q_2}{q_1},
\end{equation}
which means that any Miura panel at the bottom is always parallel to the corresponding panel at the top. 

\begin{figure}
\centering
\makebox[0pt]{\includegraphics[scale=1]{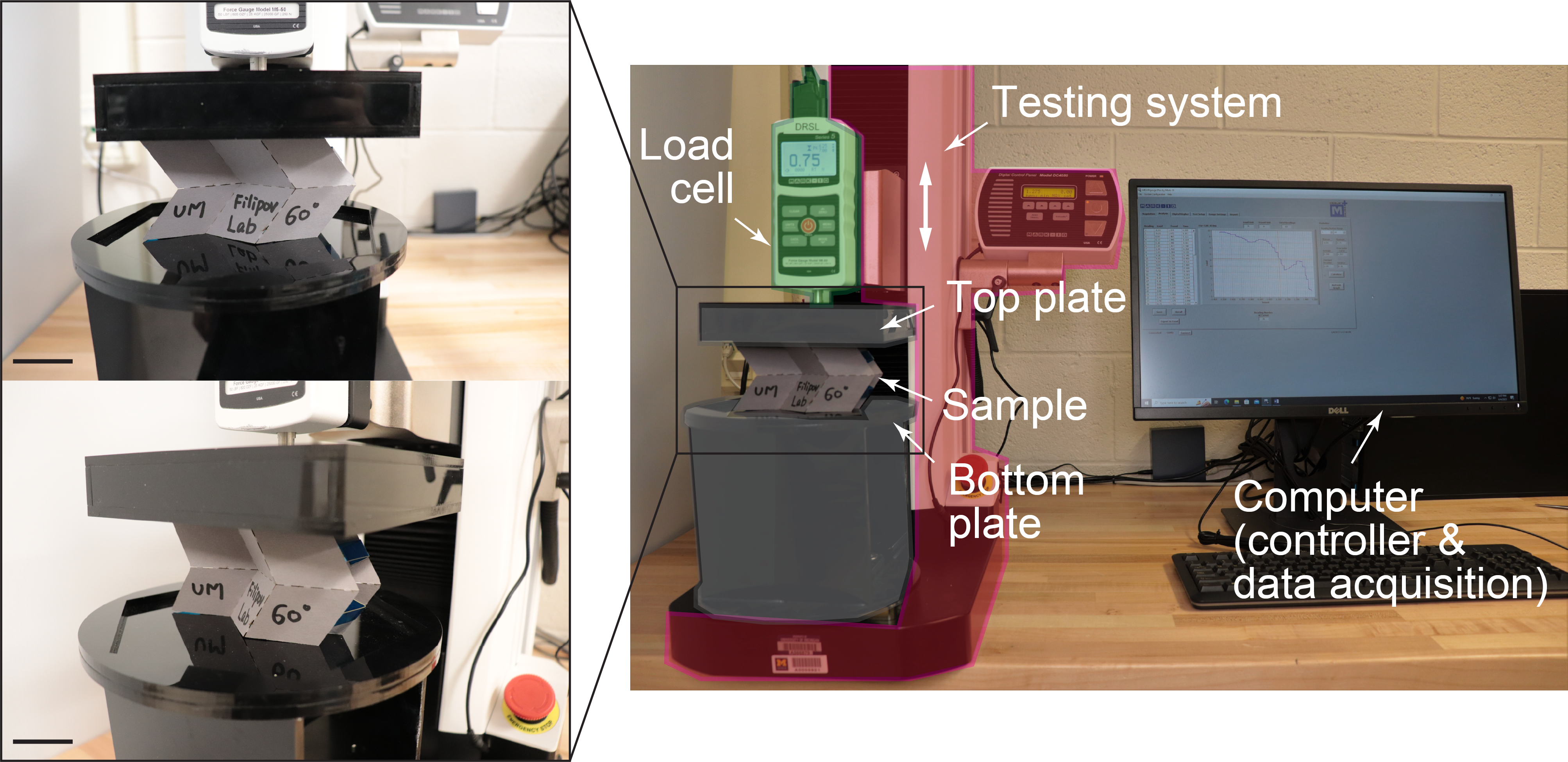}} 
\caption{Experimental set-up for testing the basic assemblies (scale bars are 4 cm).}\label{fig:TestSetUp}
\end{figure}

As mentioned earlier, the Miura-ori is a S-DOF mechanism fully controlled by the folding angle $\theta$. Two panels being parallel indicates that the bottom Miura sheet and the top Miura sheet have the same folding angle, so the top sheet is just the result of a rigid body translation of the bottom one. Also, Eq.~\ref{eqn:Geometry_2} shows that all links have the same rotation angle $\beta_i$, and thus all links on the same side of the s-s axis are parallel. This is why the number and locations of the links do not matter, as mentioned in Sec.~\ref{sec:Design}. We assume that the translation from the bottom Miura to the top Miura as $\Delta x$, $\Delta y$, and $\Delta z$---three relative displacements along three Cartesian directions, and we refer to all $\beta_i$ as just $\beta$. During rigid folding, the folded Miura sheets and the rotated links must always be symmetric about the s-s axis (the central axis). Therefore, the top Miura has no relative displacement to the bottom Miura in the $x$ direction, that is, 
\begin{equation}\label{eqn:Condition_1}
\Delta x = 0.
\end{equation}
Now, Eq.~\ref{eqn:Condition_1} is our compatibility condition, yet it is not an explicit relationship between the folding angle $\theta$ and the link rotation angle $\beta$, so next we express the $\Delta x$ using the geometric parameters defined in Fig.~\ref{fig:Geometry}. 

The $\Delta x$ is also the relative displacement from link vertex $o_1$ to link vertex $r_1$ in the $x$ direction. This displacement is not easy to obtain in the global coordinate system ($x, y, z$), but in the local coordinate system ($x_1, y_1, z_1$), one can easily see that
\begin{equation}\label{eqn:Local_Disp}
\Delta {x_1} = 0 \ , \ {\rm{ }}\Delta {y_1} = d\cos \beta \ , \ {\rm{ }}\Delta {z_1} = d\sin \beta .
\end{equation}
Therefore, we can obtain $\Delta x$ by transforming the local displacement $\Delta x_1$, $\Delta y_1$, $\Delta z_1$ to global displacement $\Delta x$, $\Delta y$, $\Delta z$. For that purpose, we obtain the rotation matrix $\bf{R}$ from the global coordinate system ($x, y, z$) to the local coordinate system ($x_1, y_1, z_1$):
\begin{equation}\label{eqn:Trans_Matrix}
\begin{array}{l}
\mathop {\bf{R}} \limits_{\{ x,y,z\}  \to \{ x_1,y_1,z_1\} } = \left( {\begin{array}{*{20}{c}}
{\cos \left\langle {{\bf{x}},{{\bf{x}}_1}} \right\rangle }&{\cos \left\langle {{\bf{x}},{{\bf{y}}_1}} \right\rangle }&{\cos \left\langle {{\bf{x}},{{\bf{z}}_1}} \right\rangle }\\
{\cos \left\langle {{\bf{y}},{{\bf{x}}_1}} \right\rangle }&{\cos \left\langle {{\bf{y}},{{\bf{y}}_1}} \right\rangle }&{\cos \left\langle {{\bf{y}},{{\bf{z}}_1}} \right\rangle }\\
{\cos \left\langle {{\bf{z}},{{\bf{x}}_1}} \right\rangle }&{\cos \left\langle {{\bf{z}},{{\bf{y}}_1}} \right\rangle }&{\cos \left\langle {{\bf{z}},{{\bf{z}}_1}} \right\rangle }
\end{array}} \right), \ \text{with}\\
\left\{ \begin{array}{l}
\cos \left\langle {{\bf{x}},{{\bf{x}}_1}} \right\rangle  = \frac{{{c_4}( - {c_1}{c_3}{c_5} + {c_6}{c_1}^2 + {c_6}{c_2}^2)}}{{\sqrt {\left( {{c_1}^2{c_4}^2 + {c_2}^2({c_4}^2 + {c_5}^2)} \right)\left( {{c_3}^2({c_4}^2 + {c_5}^2) - 2{c_1}{c_3}{c_5}{c_6} + {c_6}^2({c_1}^2 + {c_2}^2)} \right)} }},\\
\cos \left\langle {{\bf{y}},{{\bf{x}}_1}} \right\rangle  = \frac{{{c_1}{c_3}{c_4}^2 + {c_5}{c_6}{c_2}^2}}{{\sqrt {({c_1}^2{c_4}^2 + {c_2}^2({c_4}^2 + {c_5}^2))({c_3}^2({c_4}^2 + {c_5}^2) - 2{c_1}{c_3}{c_5}{c_6} + {c_6}^2({c_1}^2 + {c_2}^2))} }},\\
\cos \left\langle {{\bf{z}},{{\bf{x}}_1}} \right\rangle  = \frac{{{c_2}({c_3}({c_4}^2 + {c_5}^2) - {c_1}{c_5}{c_6})}}{{\sqrt {\left( {{c_1}^2{c_4}^2 + {c_2}^2({c_4}^2 + {c_5}^2)} \right)\left( {{c_3}^2({c_4}^2 + {c_5}^2) - 2{c_1}{c_3}{c_5}{c_6} + {c_6}^2({c_1}^2 + {c_2}^2)} \right)} }},\\
\cos \left\langle {{\bf{x}},{{\bf{y}}_1}} \right\rangle  = \frac{{ - {c_3}{c_4}}}{{\sqrt {{c_3}^2{c_4}^2 + {c_2}^2{c_6}^2 + {{({c_1}{c_6} - {c_3}{c_5})}^2}} }},\\
\cos \left\langle {{\bf{y}},{{\bf{y}}_1}} \right\rangle  = \frac{{{c_1}{c_6} - {c_3}{c_5}}}{{\sqrt {{c_3}^2{c_4}^2 + {c_2}^2{c_6}^2 + {{({c_1}{c_6} - {c_3}{c_5})}^2}} }},\\
\cos \left\langle {{\bf{z}},{{\bf{y}}_1}} \right\rangle  = \frac{{{c_2}{c_6}}}{{\sqrt {{c_3}^2{c_4}^2 + {c_2}^2{c_6}^2 + {{({c_1}{c_6} - {c_3}{c_5})}^2}} }},\\
\cos \left\langle {{\bf{x}},{{\bf{z}}_1}} \right\rangle  = \frac{{{c_2}{c_5}}}{{\sqrt {{c_2}^2{c_5}^2 + {c_2}^2{c_4}^2 + {c_1}^2{c_4}^2} }},\\
\cos \left\langle {{\bf{y}},{{\bf{z}}_1}} \right\rangle  = \frac{{ - {c_2}{c_4}}}{{\sqrt {{c_2}^2{c_5}^2 + {c_2}^2{c_4}^2 + {c_1}^2{c_4}^2} }},\\
\cos \left\langle {{\bf{z}},{{\bf{z}}_1}} \right\rangle  = \frac{{{c_1}{c_4}}}{{\sqrt {{c_2}^2{c_5}^2 + {c_2}^2{c_4}^2 + {c_1}^2{c_4}^2} }},
\end{array} \right.
\end{array}
\end{equation}
where $\langle \boldsymbol{\cdot}, \boldsymbol{\cdot} \rangle$ represents the angle between two vectors, and $c_1 - c_6$ are six constants defined by
\begin{equation}\label{eqn:Constants}
\begin{array}{l}
{c_1} = L, \ {c_2} = a\sin \psi ,\ {c_3} = \sqrt {{L^2} + {{(a\sin \psi )}^2}} \cos \eta, \\
{c_4} =  - S,\ {c_5} = b\cos \xi ,\ {c_6} = \sqrt {{S^2} + {{(b\cos \xi )}^2}} \cos (\gamma  - \eta ), \ \text{with}\\
\left\{ \begin{array}{l}
H = a\sin \gamma \sin \theta, \\
S = b\frac{{\tan \gamma \cos \theta }}{{\sqrt {1 + {{\tan }^2}\gamma {{\cos }^2}\theta } }},\\
L = a\sqrt {1 - {{\sin }^2}\gamma {{\sin }^2}\theta }, \\
V = b\frac{1}{{\sqrt {1 + {{\tan }^2}\gamma {{\cos }^2}\theta } }},
\end{array} \right.{\rm{  }}\left\{ \begin{array}{l}
\tan \xi  = \cos \theta \tan \gamma, \\
\sin \psi  = \sin \theta \sin \gamma, \\
\cos \gamma  = \cos \xi \cos \psi. 
\end{array} \right.
\end{array}
\end{equation}
The derivation of Eq.~\ref{eqn:Trans_Matrix} is given in Sec. S1.1 of the Supplementary Material. With the rotation matrix $\bf{R}$, we have the coordinate transformation as
\begin{equation}\label{eqn:Delta_xyz}
\left( \begin{array}{l}
\Delta x\\
\Delta y\\
\Delta z
\end{array} \right) = {\bf{R}}\left( \begin{array}{l}
\Delta {x_1}\\
\Delta {y_1}\\
\Delta {z_1}
\end{array} \right) = d\left( \begin{array}{l}
\frac{{ - {c_3}{c_4}\cos \beta }}{{\sqrt {{c_3}^2{c_4}^2 + {c_2}^2{c_6}^2 + {{({c_1}{c_6} - {c_3}{c_5})}^2}} }} - \frac{{{c_2}{c_5}\sin \beta }}{{\sqrt {{c_2}^2{c_5}^2 + {c_2}^2{c_4}^2 + {c_1}^2{c_4}^2} }}\\
\frac{{({c_1}{c_6} - {c_3}{c_5})\cos \beta }}{{\sqrt {{c_3}^2{c_4}^2 + {c_2}^2{c_6}^2 + {{({c_1}{c_6} - {c_3}{c_5})}^2}} }} + \frac{{{c_2}{c_4}\sin \beta }}{{\sqrt {{c_2}^2{c_5}^2 + {c_2}^2{c_4}^2 + {c_1}^2{c_4}^2} }}\\
\frac{{{c_2}{c_6}\cos \beta }}{{\sqrt {{c_3}^2{c_4}^2 + {c_2}^2{c_6}^2 + {{({c_1}{c_6} - {c_3}{c_5})}^2}} }} - \frac{{{c_1}{c_4}\sin \beta }}{{\sqrt {{c_2}^2{c_5}^2 + {c_2}^2{c_4}^2 + {c_1}^2{c_4}^2} }}
\end{array} \right).
\end{equation}
Finally, we let $\Delta x = 0$ and get the explicit kinematic compatibility equation as
\begin{equation}\label{eqn:Final_Eq}
\left( {\frac{{ - {c_3}{c_4}}}{{\sqrt {{c_3}^2{c_4}^2 + {c_2}^2{c_6}^2 + {{({c_1}{c_6} - {c_3}{c_5})}^2}} }}} \right)\cos \beta  = \left( {\frac{{{c_2}{c_5}}}{{\sqrt {{c_2}^2{c_5}^2 + {c_2}^2{c_4}^2 + {c_1}^2{c_4}^2} }}} \right)\sin \beta.
\end{equation}
Eq.~\ref{eqn:Final_Eq} is an analytical, quantitative relationship between the Miura folding angle $\theta$ and the link rotation angle $\beta$. We can solve for the value of $\beta$ once we substitute a value of $\theta$ into $c_1 - c_6$, and in this way, we obtain the geometry of the entire basic assembly. 

Taking possible singularities of origami configurations and possible contact between origami panels into consideration \citep{zhu2019efficient}, we obtain four different solutions to Eq.~\ref{eqn:Final_Eq} that mean four different types of $\beta$--$\theta$ relations, and those relations correspond to different folding paths and mechanical properties of our basic assemblies. We will discuss these different folding paths in detail in Sec.~\ref{sec:Properties}. 

Note that the procedure of deriving the compatibility equation here is generalized. Even if we include more geometric parameters into the basic assembly design (for instance, an arbitrary initial link rotation angle $\beta_0$), the compatibility condition in Eq.~\ref{eqn:Condition_1} still holds and a similar governing kinematic equation can be derived through parameterization of the compatibility condition. With specific values assigned to the parameters of an extended basic assembly, classic 3D dense cellular origami designs such as the stacked Miura sandwich \citep{schenk2013geometry} can be obtained. More details can be found in Sec. S1.2 of the Supplementary Material. As more complicated designs do not bring new folding paths while making the fabrication more difficult, we only focus on our basic assembly as described in Sec.~\ref{sec:Design}. 


\subsection{Modeling of the elastic-folding mechanics}\label{sec:Mechanics}
In this sub-section, we model our basic assemblies from a mechanics perspective to explore elastic deformation to validate the kinematic simulations presented in Sec.~\ref{sec:Kinematics}, and to explore any elastic deformation phenomena that may arise during folding of the systems. We use a reduced-order finite element method (FEM) where we treat our basic assembly as a truss structure and model its elastic-folding mechanics with a structural engineering technique. In this reduced-order FEM, we only use two types of elements---bars and torsional springs (hinges). The basic idea of this \textit{bar} \& \textit{hinge model} is to characterize the in-plane stretching and shearing with bars, out-of-plane bending and twisting with bending hinges, and folding along creases with folding hinges \citep{filipov2017bar}. 

The bar \& hinge model is usually for 2D origami sheets, while in this study we generalize it to the 3D multi-layered origami sheet system. For our basic assemblies, all folding creases naturally become the folding hinges; origami panels are prone to bending along the diagonals (especially the shorter diagonals), so the line segments on the panels become the bending hinges; the bars are placed such that the nodes on the links are properly connected to the nodes on the sheets. We present the bar \& hinge mesh for a basic assembly in Fig.~\ref{fig:Geometry}(g). This mesh has 60 nodes (180 DOFs) and 64 triangular panels in total. The initial nodal coordinates are defined based on the design parameters of the basic assembly. 

We next add support constraints and loads to the mesh (see Fig.~\ref{fig:Geometry}(g)). We apply six basic constraints on three nodes to eliminate rigid body movement of the basic assembly (including three translations and three rotations). The folding load is placed on eight nodes on the outer edge, and all eight nodal forces have the same magnitude and direction (along `$-x$'). In addition to a folding load, we apply a perturbing load\footnote{We use \textit{load perturbation} instead of \textit{geometric perturbation} (which is another strategy to initiate the origami folding by slightly pre-folding the creases) because we do not know the exact geometry of the folded basic assembly in advance (or even whether the basic assembly is foldable).} on four nodes to initiate the origami folding because we start our mechanics simulation with the completely flat state in this study. The directions of four perturbing forces follow the directions of nodal displacements of a folded Miura-ori, and the magnitude of the perturbing load is set to 1/100 of that of the folding load. We measure the displacement along the loading direction (`$-x$' direction) which is denoted as $\Delta$. 

We define elastic behavior of the bar \& hinge model based on the material properties of the physically fabricated assemblies. The paper sheets and plastic strips have Young's moduli and Poisson's ratios of $E_1=1.5$ GPa (paper), $E_2=4$ GPa (plastic), and $\nu_1=\nu_2=0.3$. The thicknesses of the two materials are $t_1=0.8$ mm and $t_2=0.6$ mm. According to the experimental constituent relations of folded thin sheets \citep{wo2023stiffening,filipov2017bar,woodruff2020bar}, we can determine the values of the equivalent bar \& hinge parameters, including the folding stiffness $K_F$, bending stiffness $K_B$, and stretching stiffness $K_S$, using the material properties and the geometry of the triangular panels. The folding stiffness $K_F$ and bending stiffness $K_B$ are given by
\begin{equation}\label{eqn:FBStiffness}
{K_F} = \frac{{{L_F}}}{{{L^*}}}\frac{{E{t^3}}}{{12(1 - {\nu ^2})}}, \ \ {K_B} = {\left( {\frac{{{L_B}}}{t}} \right)^{1/3}}\frac{{E{t^3}}}{{12(1 - {\mu ^2})}}
\end{equation}
where $L_F$ and $L_B$ are the lengths of the folding hinge and bending hinge, and $L^*$ is the characteristic length scale, which is typically proportional to the material thickness (in this work, we set $L^*=600t$ to represent creases that are much more flexible than the bending hinges) \citep{lechenault2014mechanical}. The stretching stiffness is given by
\begin{equation}\label{eqn:BarArea}
A = \frac{{0.36St}}{{C(1 - \nu )}}{\left( {\frac{h}{L_S}} \right)^{1/3}}, \ \ {\rm{ }}{K_S} = \frac{{EA}}{L_S},
\end{equation}
where $A$ is the cross-sectional area of the bar, $L_S$ is the length of the bar, $S$ and $C$ are the area and perimeter of the triangular panel, and $h$ is the height of the triangular panel associated with the bar. If a bar is shared by more than one triangle, its stretching stiffness is calculated by summing the stiffness contributed by each panel (see Sec. S1.3 of the Supplementary Material for an example of how we determine the values of bar \& hinge parameters based on the material and geometric properties). 

After meshing the basic assembly, defining the constraints, loads, and material constants, we solve for a displaced equilibrium state. In this study, we use the \textit{modified generalized displacement control method} (MGDCM) as the solver \citep{leon2014effect}. As a variant of the arc-length method, the MGDCM is an efficient algorithm for tracking complicated nonlinear equilibrium paths. The load step size in the MGDCM is adaptively adjusted, allowing for reasonable accuracy and faster convergence. The tangent stiffness matrix needed for MGDCM iterations is formulated by a potential energy approach \citep{liu2017nonlinear}. The total potential energy $\Pi$ of our origami systems can be computed as:
\begin{equation}\label{eqn:StrainEnergy}
\begin{array}{l}
\begin{split}
\Pi &= U_{\mathrm{folding}} + U_{\mathrm{bending}} + U_{\mathrm{stretching}} + V_{\mathrm{external}}\\
 &= \sum\limits_{n = 1}^N {\frac{1}{2}{K_{F,n}}{{\left( {{\varphi _{F,n}} - \varphi _{F,n}^0} \right)}^2}}  + \sum\limits_{m = 1}^M {\frac{1}{2}{K_{B,m}}{{\left( {{\varphi _{B,m}} - \varphi _{B,m}^0} \right)}^2}}  + \sum\limits_{q = 1}^Q {\frac{1}{2}{K_{S,q}}{{\left( {{L_{S,q}} - L_{S,q}^0} \right)}^2}} + {{\bf{f}}^T}{\bf{u}},
\end{split}
\end{array}
\end{equation}
where $U_{\mathrm{folding}}$, $U_{\mathrm{bending}}$, and $U_{\mathrm{stretching}}$ represent the internal folding energy, bending energy, and stretching energy computed based on the rotation of the folding hinges, rotation of the bending hinges, and deformation of the bars; while $V_{\mathrm{external}}$ represents the potential of the external load. Here, $\varphi_{F}^0$ and $\varphi_{B}^0$ represent the initial values of the folding and bending angles respectively, $L_{S}^0$ represent the initial bar lengths when the origami is unfolded, and $\bf{f}$ and $\bf{u}$ are the nodal load and displacement vectors. When a perturbation load exists, it is included in the nodal load vector $\bf{f}$.


\subsection{Experimental testing}\label{sec:Setup}
We run experimental tests to verify our simulations using a Mark-10\textsuperscript{®} ESM 1500 single-column tabletop testing system with a 250 N load cell (see Fig.~\ref{fig:TestSetUp}). Our prototypes are placed between a top and a bottom plate, both made of acrylic. The assembly is free to slide with respect to the two plates, and we did not apply any constraints because the deformation of the basic assembly can be complex and any additional boundary conditions could over-constrain the system. We use a displacement control, and a loading rate of 60 mm/min is used for all tests in this study. We record the displacement and the compression force at the same time with a sampling rate of 20 Hz. In the elastic mechanics modeling in Sec.~\ref{sec:Mechanics}, we initiate the origami folding using load perturbation; while in the experimental testing, geometric perturbation by slightly pre-folding the creases is much easier to implement and is thus used in all experimental tests. 

\section{Kinematic and mechanical properties}\label{sec:Properties}
In this section, we use the modeling and testing approach described in Sec.~\ref{sec:Modeling} to explore the kinematic and mechanical properties of the basic assembly, and thus to find out how different ways of connecting the sheets and the linkages lead to different categories of multi-layered spaced origami. The magnitude of the folding load is set to 1 and the initial load factor in the MGDCM iteration \citep{leon2014effect} is set to 0.2 for all nonlinear mechanics simulations. 

\begin{figure}
\centering
\makebox[0pt]{\includegraphics[scale=1]{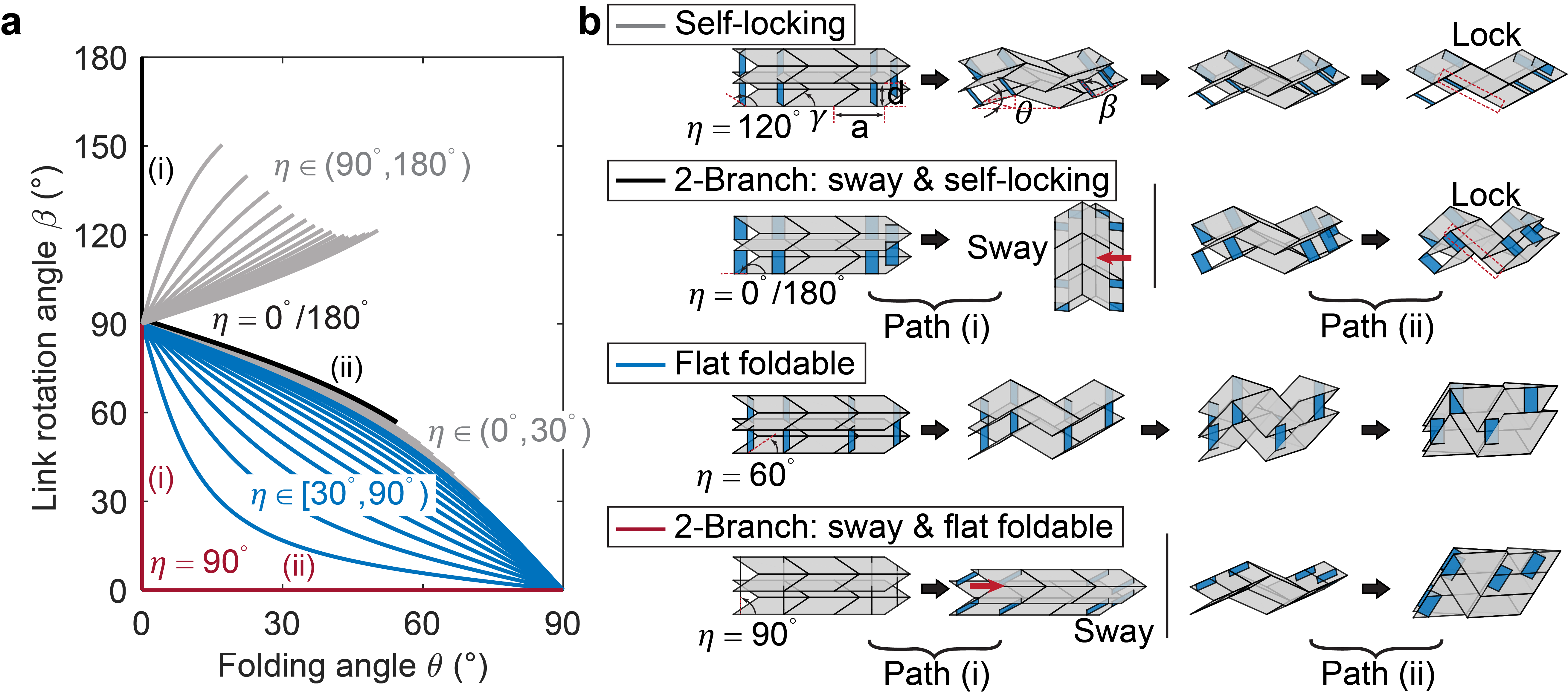}} 
\caption{Rigid-folding kinematics of multi-layered spaced origami. (a) The relationship between the link rotation angle $\beta$ and the folding angle $\theta$. Each line represents a structure with a different link orientation angle $\eta$ in the range [0, $\pi$]; (b) Visualization of the four types of folding kinematics which are represented as different line colors in (a).}\label{fig:Kinematics}
\end{figure}

\subsection{Rigid folding paths}\label{sec:RigidFoldingPath}
In this section, we derive all the possible solutions to the kinematic compatibility equation (Eq.~\ref{eqn:Final_Eq} that we derived earlier in Sec.~\ref{sec:Kinematics}). Based on these solutions we compute the rigid folding paths of the multi-layered spaced origami. The two denominators in the coefficients of `$\cos\beta$' and `$\sin\beta$' in Eq.~\ref{eqn:Final_Eq} are always positive, so the properties of the solutions only depend on `$c_3 c_4$' and `$c_2 c_5$', the two numerators. We have five different cases: 
\begin{itemize}
    \item The General Case: $c_3c_4\neq0$, $c_2c_5\neq0$. The $\beta$--$\theta$ relation in this case is given by
\begin{equation}\label{eqn:GeneralSolution}
\begin{array}{l}
\beta  = \arctan \left( {\frac{{ - {c_3}(\theta ){c_4}(\theta )\sqrt {{c_2}^2(\theta ){c_5}^2(\theta ) + {c_2}^2(\theta ){c_4}^2(\theta ) + {c_1}^2(\theta ){c_4}^2(\theta )} }}{{{c_2}(\theta ){c_5}(\theta )\sqrt {{c_3}^2(\theta ){c_4}^2(\theta ) + {c_2}^2(\theta ){c_6}^2(\theta ) + {{({c_1}(\theta ){c_6}(\theta ) - {c_3}(\theta ){c_5}(\theta ))}^2}} }}} \right),\\
\text{with} \ \theta  \in [0,{\rm{ }}\pi /2), \ \beta  \in [0,{\rm{ }}\pi ],
\end{array}
\end{equation}
where each $c_i(\theta)$, $i=1,2,\cdots,6$ represents a function of $\theta$ (see Eq.~\ref{eqn:Constants}). The value of the link rotation angle $\beta$ is uniquely determined by the value of the Miura folding angle $\theta$ in this case. 
    \item Special Case 1 (\textit{Perpendicular link orientation}): $c_3c_4=0$, $c_2c_5\neq0$. This case occurs when $\eta=\frac{\pi}{2}$ and $\theta\neq0$, which indicates any folded state of a basic assembly with the links oriented perpendicularly to the edge of the sheets. The link rotation angle $\beta$ can be 0 or $\pi$ in this case, but the only valid solution is $\beta=0$ because penetration exists between origami panels when $\beta=\pi$ (as we will explain further when discussing Special Case 4). 
    \item Special Case 2 (\textit{Initial unfolded states}): $c_3c_4\neq0$, $c_2c_5=0$. This case occurs when $\eta\neq\frac{\pi}{2}$ and $\theta=0$, which indicates the initial unfolded state of a basic assembly with links that are not perpendicularly oriented to the edge of the sheets. The link rotation angle $\beta$ should be $\frac{\pi}{2}$ in this case. This result matches with our design of basic assemblies where the two Miura sheets are parallel and the links stand upright in the initial configuration (see Fig.~\ref{fig:Geometry}(a)). This special case is included in the general solution Eq.~\ref{eqn:GeneralSolution} as $\arctan(\infty)=\frac{\pi}{2}$. 
    \item Special Case 3 (\textit{Singularity}): $c_3c_4=0$, $c_2c_5=0$. This case occurs when $\eta=\frac{\pi}{2}$ and $\theta=0$, which means the initial unfolded state of a basic assembly when the links are oriented perpendicular to the edge of the sheets. The link rotation angle $\beta$ can be any value between 0 and $\pi$ in this case, leading to a singular configuration where $\beta$ is not uniquely determined by the folding angle $\theta$. The other similar singularity occurs when $\eta=0 \ \text{or} \ \pi$ and $\theta=0$ (which represents the initial unfolded state when the links are oriented parallel to the edge of the sheets)\footnote{When $\eta=0\ \text{or }\pi$ , $\theta=0$, and $\beta\neq\frac{\pi}{2}$, the kinematic compatibility condition `$\Delta x=0$' no longer holds although the structure is indeed kinematically compatible. This is the only case where that happens because this is the only case where the symmetry about the central s-s axis is broken (see Fig.~\ref{fig:Geometry}(c)).}. In short, the link rotation angle $\beta$ can be any value when the link orientation angle $\eta = 0 \ \text{or }\pi \ \text{or} \ \frac{\pi}{2}$ and the Miura folding angle $\theta=0$. 
    \item Special Case 4 (\textit{Contact}): $d({\rm{Panel}}{_i},{\rm{Panel}}{_j}) \le 0$, where $d$ is the distance between any two origami panels. The kinematic compatibility equation checks the rigidity of our basic assembly as a mechanism made of bars and hinges, but this strategy fails to detect possible contact between origami panels \citep{zhu2019efficient}. We can derive a mathematical condition for contact (see Sec. S2.1 of the Supplementary Material for explanation):
\begin{equation}\label{eqn:Condition_Contact}
\Delta z \le \left\{ \begin{array}{l}
 - \frac{H}{L}\Delta y + Hn, \ \ {\rm{if}} \ n = {\rm{ceil}}(\Delta y/L) = 2k,{\rm{  }}k =  \cdots , - 2, - 1,0,1,2, \cdots \\
\frac{H}{L}\Delta y - H(n - 1),\ \ {\rm{if}} \ n = {\rm{ceil}}(\Delta y/L) = 2k - 1,{\rm{  }}k =  \cdots , - 2, - 1,0,1,2, \cdots 
\end{array} \right.,
\end{equation}
where $\Delta y$ and $\Delta z$ represent the relative displacement of the top Miura sheet with respect to the bottom Miura sheet, which are calculated by Eq.~\ref{eqn:Delta_xyz}; $H$ and $L$ are the dimensions of a Miura-ori unit (see Fig.~\ref{fig:Geometry}(b)); the $\mathrm{ceil}(x)$ function computes the smallest integer that is greater than or equal to $x$. The existence of penetration is why we rule out the `$\beta=\pi$' solution in Special Case 1. 
    \end{itemize}

After considering all general and special cases, we compute the solutions to Eq.~\ref{eqn:Final_Eq} and thus obtain all rigid folding paths of a basic assembly with its geometry defined by $a=b=4$ cm, $\gamma=60^{\circ}$, $w=1$ cm, and $d=2$ cm. The link orientation angle $\eta$ is set to different values within [0, $\pi$], and the computed $\beta$--$\theta$ relations are given in Fig.~\ref{fig:Kinematics}. Four types of rigid folding paths exist: 1) the \textit{blue cluster} ($\eta\in[30^{\circ},90^{\circ})$) represents a \textit{flat foldable} kinematic motion. Each solution is a smooth curve from the initial state $\theta=0^{\circ},\beta=90^{\circ}$ to the final flat state $\theta=90^{\circ},\beta=90^{\circ}$, which means a continuous, rigid folding process of the origami; 2) the \textit{gray clusters} ($\eta\in(0^{\circ},30^{\circ})\ \text{or} \ (90^{\circ},180^{\circ})$) represent a kinematic path with \textit{self-locking}. Each solution starts with the initial configuration $\theta=0^{\circ},\beta=90^{\circ}$ and progresses smoothly but stops at a certain folding angle. Panel contact occurs prior to the flat state and the origami cannot be folded further; 3) the \textit{black lines} ($\eta = 0^{\circ} \ \text{or }180^{\circ}$) represent a \textit{double-branch} kinematic motion. The first branch (Path (i)) at $\theta=0^{\circ}$ is a vertical line because the initial unfolded state is a singular configuration, leading to a \textit{sway motion}. The second branch (Path (ii)) associated with $\theta>0^{\circ}$ is a kinematic path with self-locking prior to flat folding; 4) the \textit{red lines} ($\eta = 90^{\circ}$) represent the other \textit{double-branch} kinematic motion. Similarly, the first branch (Path (i)) at $\theta=0^{\circ}$ represents a sway motion, while the second branch (Path (ii)) associated with $\theta>0^{\circ}$ represents a flat foldable motion. 


\subsection{Elastic folding paths}\label{sec:ElasticFoldingPath}
\begin{figure}[!htb] 
\centering
\makebox[0pt]{\includegraphics[scale=1]{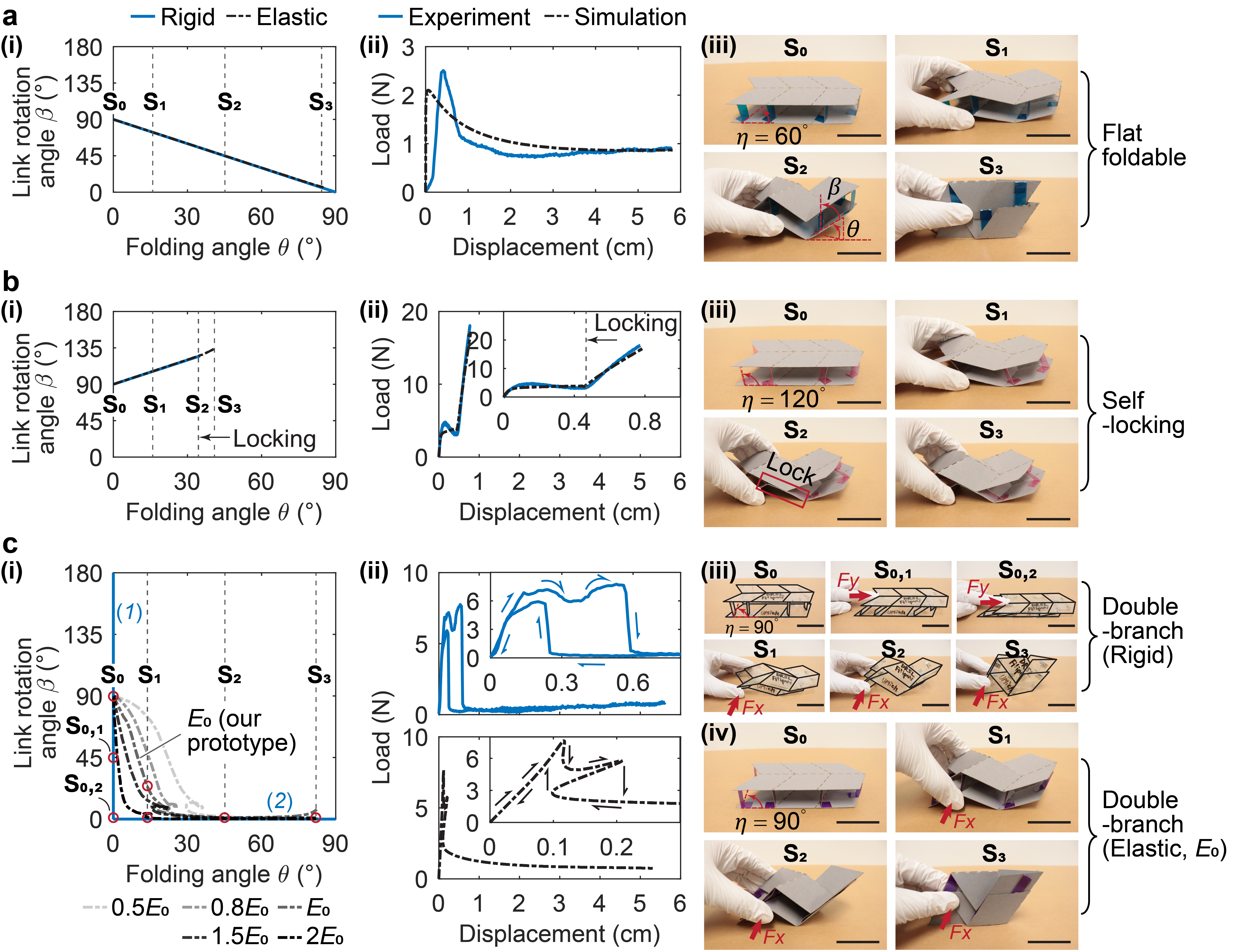}} 
\caption{Comparison between the rigid and elastic folding paths for three different folding scenarios including (a) flat foldable $\eta=60^{\circ}$, (b) self-locking $\eta=120^{\circ}$, and (c) double-branch $\eta=90^{\circ}$. In these figures, (i) show the rigid and elastic folding paths, (ii) show the simulated and experimental load--displacement curves, and (iii--iv) show the folding processes of physical prototypes. The flat foldable and self-locking cases behave similar with elastic or rigid folding assumptions, while the behavior of the double-branch folding depends on the elastic modulus. See the text for more details about (c). All scale bars are 4 cm.} \label{fig:FoldingPaths}
\end{figure}

We have identified all possible rigid folding paths of the multi-layered spaced origami, but do these structures behave the same when they are made of non-rigid deformable sheets? To answer that question, we obtain the elastic folding paths using the bar \& hinge simulation introduced in Sec.~\ref{sec:Mechanics} and we compare the two types of paths in Fig.~\ref{fig:FoldingPaths}. We choose link orientation angles of $\eta=60^{\circ}$, $120^{\circ}$, and $90^{\circ}$ to explore as typical examples of the (a) flat foldable, (b) self-locking, and (c) double-branch kinematic motions, respectively. For the flat foldable and self-locking types of kinematic motions, the analytical rigid folding motion and the simulated elastic folding exhibit no observable differences (see Fig.~\ref{fig:FoldingPaths}(a) and (b)). The effectiveness of our bar \& hinge simulation approach is demonstrated by the excellent match between the simulated and experimental load--displacement curves in Fig.~\ref{fig:FoldingPaths}(a, ii) and (b, ii). The self-locking property leads to a graded stiffness (see Fig.~\ref{fig:FoldingPaths}(b, ii)), which has various engineering applications such as impact resistance \citep{wo2022locking,ma2018origami,wen2021stacked}. 

The behavior of the double-branch kinematic motion, however, is distinctly different in terms of kinematics and mechanics. In rigid folding kinematics, the origami configuration has to pass through the intersection ($\theta=0^{\circ}$, $\beta=0^{\circ}$) for the system to switch from Path (1) to Path (2) (see Fig.~\ref{fig:FoldingPaths}(c, i)). This kinematic folding process is visualized using a rigid prototype made of thick (1.5mm) acrylic plates and nearly frictionless steel hinges (see Fig.~\ref{fig:FoldingPaths}(c, iii)). While the rigid prototype experiences almost no elastic deformation and folds according to the rigid kinematics, this is not the case when we fold an elastic prototype. As Fig.~\ref{fig:FoldingPaths}(c, i) shows, the flexibility of the materials enables a smoother transition from path (1) to path (2). The transition state S1 in Fig.~\ref{fig:FoldingPaths}(c, iv) is a configuration that is not possible for rigid origami. In other words, the flexibility of the thin sheets simplifies our actuation so that only $F_x$, instead of both $F_x$ and $F_y$, is needed for origami folding. The two transition states $\mathrm{S}_1$ and $\mathrm{S}_2$ correspond to two strain energy minima meaning two locally stable states (as shown in Fig. S5(c, i) of the Supplementary Material). These transitions also cause two sudden drops in the forward path of the load--displacement curve (see Fig. S5(c, ii)), which makes the origami system exhibit a snap-through-like behavior. However, the states $\mathrm{S}_1$ and $\mathrm{S}_2$ are both associated with a non-zero external load, so the system does not have `load-free multi-stability' here. In our exploration of the multi-layered spaced origami, we have not identified multistable behaviors to occur, however such characteristics may be present with other stiffness parameters or loading of the structures.

Despite the smooth transition observed here, the elastic folding path approaches the rigid folding path as we increase Young's modulus of the materials, and vice versa, as Fig.~\ref{fig:FoldingPaths}(c, i) shows ($E_0$ denotes the Young's modulus of the materials used to fabricate the elastic prototype in Fig.~\ref{fig:FoldingPaths}(c, iv), while $0.5E_0,0.8E_0,1.5E_0$, and $2E_0$ represent other materials with lower or higher Young's moduli). The additional sway DOF of the double-branch kinematic motion makes the spacing between the sheets adjustable. This motion which is demonstrated in the top of Fig.~\ref{fig:FoldingPaths}(c, iii) can be harnessed to design functional devices with tunable properties. We will give an example later in Sec.~\ref{sec:Applications}. All the above folding behaviors can also be interpreted from a strain energy perspective and more details can be found in Sec. S2.2 of the Supplementary Material.

\subsection{Rigid foldability, flat foldability, and packing ratio}\label{sec:Foldability}
Rigid foldability and flat foldability are the most talked-about properties in origami design. Rigid foldability allows engineers to construct origami structures with rigid, durable materials, while flat foldability makes origami structures as compact as possible when folded. Fig.~\ref{fig:FoldingPaths}(a) demonstrates the folding kinematics and mechanics of the multi-layered spaced origami corresponding to the blue lines in Fig.~\ref{fig:Kinematics}(a) (the flat foldable type). The change of volume of the basic assembly observed in Fig.~\ref{fig:FoldingPaths}(a, iii) indicates that the designed origami has a large \textit{packing ratio}, which is defined as the volume of the unfolded origami/the volume of the folded or stowed origami. In this work, we define the volume of the origami as the volume of the circumscribed cuboid of the entire origami system. In the stowed configuration, we calculate the volume when the folding angle $\theta=85^{\circ}$. 

\begin{figure}[!htb] 
\centering
\makebox[0pt]{\includegraphics[scale=1]{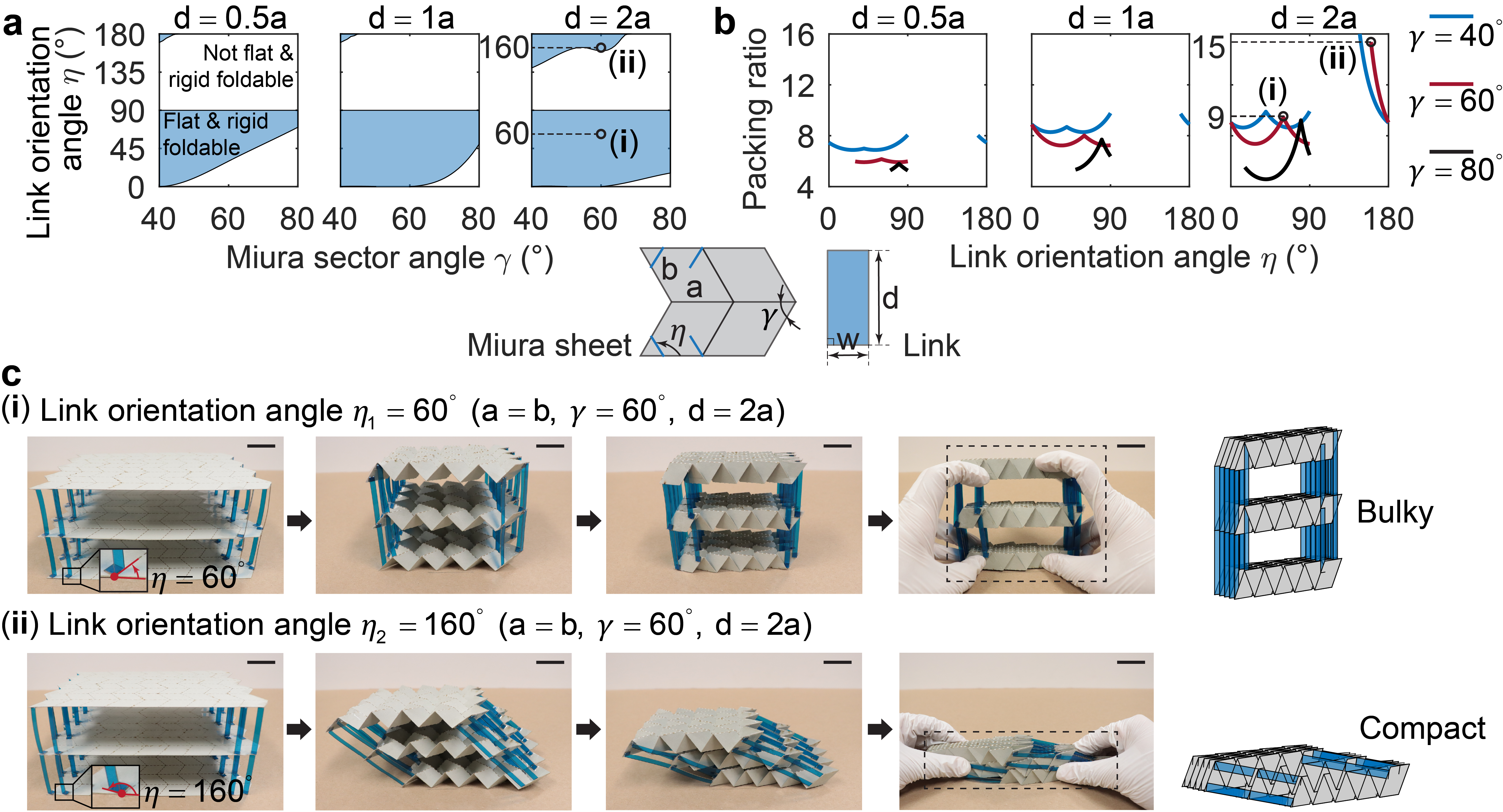}} 
\caption{Design space of rigid, flat foldable multi-layered spaced origami and the corresponding packing ratios. (a) The $\eta$--$\gamma$ design space for different sheet spacings defined by link lengths $d$ of $0.5a$, $a$, and $2a$. The blue regions indicate the available design space where the system is rigid and flat foldable; (b) Computed packing ratios. The curves are shown for $\gamma$ angles of $40^{\circ}$, $60^{\circ}$, and $80^{\circ}$, while the $\eta$ angle ($x$ axis) varies within the corresponding design space in (a); (c) An example of how the link orientation angle $\eta$ affects the packing ratio when the sheet spacing is relatively large ($d=2a$). The two designs are associated with the points marked (i) and (ii) in (a) and (b). The two origami structures are both $4\times3\times2$ blocks with the only difference being $\eta_1=60^{\circ}$ and $\eta_2=160^{\circ}$ (all scale bars are 2 cm).} \label{fig:PackRatio}
\end{figure}

Fig.~\ref{fig:PackRatio} shows how the Miura sector angle $\gamma$, the link orientation angle $\eta$, and the length of the links $d$ influence the rigid and flat foldable design space and the corresponding packing ratio of the multi-layered origami. When $d$ is relatively small ($d=0.5a$), only around $\frac{1}{3}$ of the $\eta$--$\gamma$ space is available for rigid, flat foldable design. The design space is limited by the self-contact kinematic path shown in Fig.~\ref{fig:FoldingPaths}(b). Because $d$ is also the spacing between the deployed Miura sheets, a small $d$ also leads to a more compact configuration when the multi-layered origami is stowed. When $d$ is small, the other design parameters have little influence on the packing ratio (see Fig.~\ref{fig:PackRatio}(b)). As $d$ increases, the design space for rigid and flat foldability expands because a larger spacing between the sheets results in less interference during folding. For example, when $d=2a$, around $\frac{3}{4}$ of the $\eta$--$\gamma$ space is available. However, when $d$ is larger, the packing ratio changes substantially when the other design variables are changed. The link orientation angle $\eta$ determines whether we can have an `optimal packing'. For example, when the Miura sector angle $\gamma$ is $60^{\circ}$, setting the link orientation angle $\eta$ to $60^{\circ}$ results in a structure with a packing ratio of 9 (marked with (i) in Fig.~\ref{fig:PackRatio}(a, b)). However, almost twice this value can be obtained if we set $\eta$ to $160^{\circ}$ (marked with (ii) in Fig.~\ref{fig:PackRatio}(a, b)). The two folding processes are shown in Fig.~\ref{fig:PackRatio}(c), and while the deployed structures look similar the stowed geometries are markedly different. When the inner spacing in a multi-layered structure is relatively large, we can strategically select the other design parameters, especially the link orientation angle $\eta$, to optimize the packaging ratio. As such, we can design multi-layered spaced origami to achieve compact packing while maintaining rigid foldability and flat foldability \citep{leanza2022hexagonal,lu2023easy}.

\subsection{Directional stiffness}\label{sec:Stiffness}
\begin{figure}[!htb] 
\centering
\makebox[0pt]{\includegraphics[scale=1]{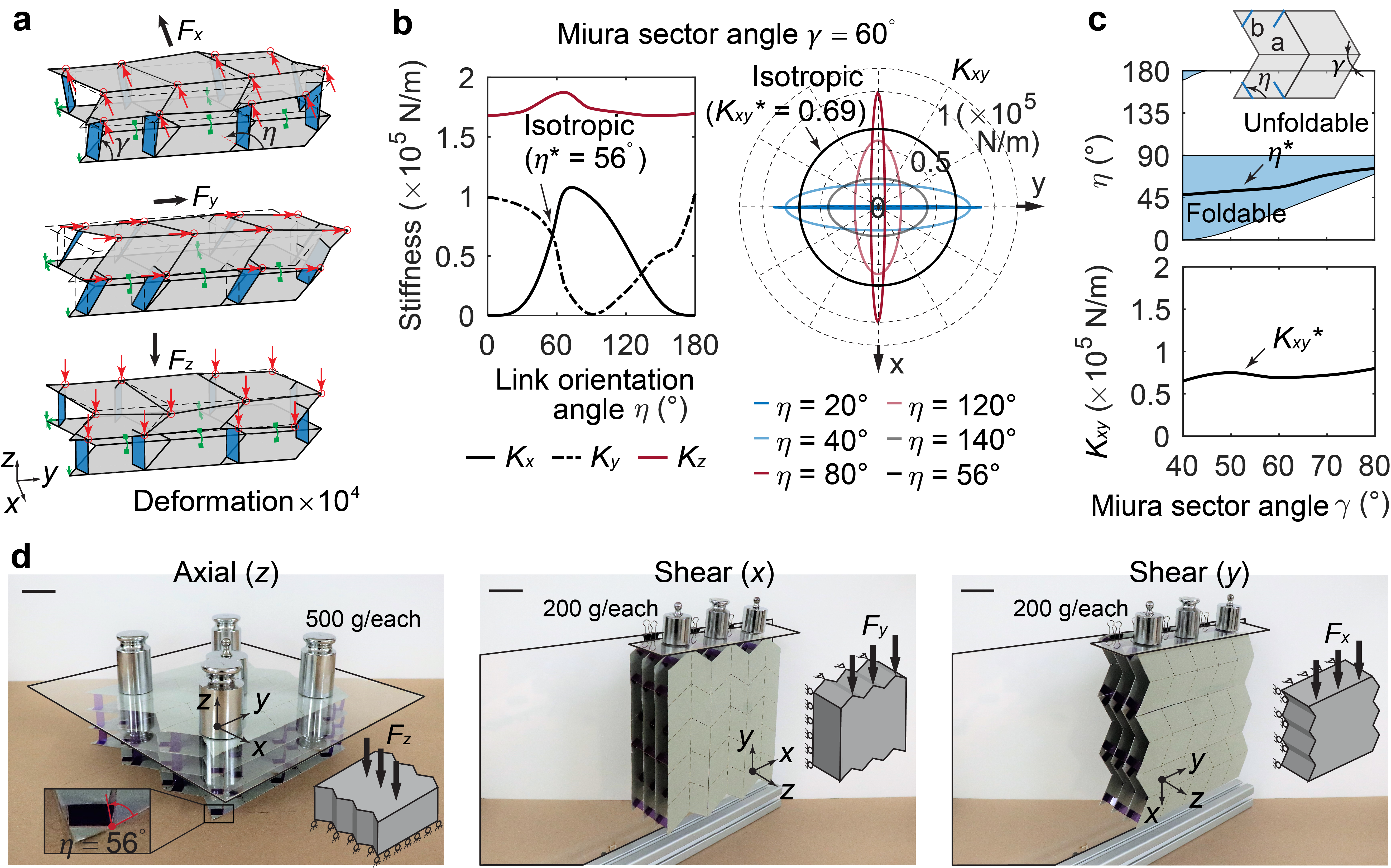}} 
\caption{Directional stiffness characteristics of the basic assembly. (a) Deformation of the basic assembly where the bottom Miura sheet is locked and distributed loads $F_x$, $F_y$, and $F_z$ are applied on the top Miura sheet in the three Cartesian directions (in this example, the link orientation is $\eta=120^{\circ}$, and deformations are scaled by $10^4$); (b) Left: Stiffness of the basic assembly in the three Cartesian directions $K_x$, $K_y$, and $K_z$ versus the link orientation angle $\eta$; Right: A polar plot representing the stiffness of the structure when loaded in different directions in the $x$-$y$ plane (the stiffness $K_{xy}$). Curves are shown for different link orientation angles $\eta$, and the case with an isotropic stiffness $K_{xy}^*$ is identified; (c) The isotropic link orientation angle $\eta^*$ and the corresponding stiffness $K_{xy}^*$ for different Miura sector angles $\gamma$; (d) A physical prototype that demonstrates the load-bearing capacity by isotropic design (the sample is a $2\times3\times3$ block, and the link orientation angle is $56^{\circ}$, the value of $\eta^*$ when $\gamma=60^{\circ}$; As we expect, the structure has a high relative stiffness in all three Cartesian directions, and the behavior appears uniform for in-plane loading. All scale bars are 4 cm).}\label{fig:Stiffness}
\end{figure}

Designing origami structures that have reasonable stiffness and prevent unwanted deformations has been an ongoing challenge in engineering because origami initially has kinematic folding modes and is made of thin sheets that are prone to bending and buckling \citep{filipov2015origami,jamalimehr2022rigidly,overvelde2016three}. In this sub-section, we evaluate the stiffness of our basic assemblies along different directions. We first set up the constraints (boundary conditions) and loads in the bar \& hinge mechanics simulation. We first apply six constrains on the base of the structure to eliminate the rigid body movement of the basic assembly in 3D space (see Fig.~\ref{fig:Stiffness}(a)). Origami structures are usually locked in place after deployment \citep{reis2015transforming,meloni2021engineering}, so for our basic assembly, we lock three of the folding creases on the bottom Miura sheet by making their folding stiffness large (see the `green locks' on the folding creases in Fig.~\ref{fig:Stiffness}(a)). With these constraints, the rigid folding mode and rigid body motions are eliminated, yet the entire structure is not overly constrained. We apply a distributed load to the basic assembly by placing forces of the same magnitude and same direction onto 12 nodes in the top Miura sheet. Fig.~\ref{fig:Stiffness}(a) shows how the constraints and loads are applied and the resulting deformed shapes when the basic assembly is loaded along three Cartesian axes (the deformation is amplified $10^4$ times). 

Here, we only compute the linear stiffness under the assumption of small deformations by directly evaluating the stiffness matrix $\mathbf{K}$. We calculate all nodal displacements using $\mathbf{F}=\mathbf{K}\mathbf{u}$ where $\mathbf{F}$ is the nodal force vector and $\mathbf{u}$ is the nodal displacement vector. Finally, we obtain a scalar representative stiffness $K$ based on the definition $K=F/\Delta$, where $F$ is the total magnitude of the applied loads and $\Delta$ is the average nodal displacement associated with all 12 nodes on the top Miura sheet. 

Figure~\ref{fig:Stiffness}(b, the left panel) shows how the directional stiffness of the basic assembly varies with the link orientation angle (with $\gamma=60^{\circ}$, $a=b$, and $d=0.5a$). In general, the double-layered basic assembly with the bottom Miura sheet locked mimics the behavior of a cantilever beam oriented along the $z$ axis: i) the `axial stiffness' $K_z$ is always higher than the `bending stiffness' $K_x$ and $K_y$, and it remains relatively consistent with respect to the link orientation; and ii) the bending stiffness in the two orthogonal directions $K_x$ and $K_y$ change with the link orientation with opposing trends. When the link orientation angle $\eta$ approaches $90^{\circ}$, the basic assembly is stiff in the $x$ direction while flexible in the $y$ direction---the opposite is true when $\eta$ approaches $0^{\circ}$ or $180^{\circ}$. Similar to a cantilever with a rectangular cross-section, the basic assembly with $\eta=0^{\circ}$ or $90^{\circ}$ or $180^{\circ}$ has a high bending moment of inertia along one Cartesian axis and a low stiffness along the other axis within the $x$-$y$ plane. While the behavior is similar to cantilevers (and an individual link), the stiffness of the entire multi-layered basic assembly is not symmetric when the links are rotated (that is, $K_x$ and $K_y$ curves are not symmetric about $\eta=90^{\circ}$ in Fig.~\ref{fig:Stiffness}(b, the left panel)). The loss of symmetry is caused by the additional sway deformation mode of the origami at $\eta=90^{\circ}$ that we discussed in Sec.~\ref{sec:RigidFoldingPath}. Further explanations are provided in Sec. S3.1 of the Supplementary Material. 

We then rotate the loads in the $x$-$y$ plane and obtain the stiffness of the basic assembly for arbitrary directions of in-plane loading ($K_{xy}$ shown as a polar plot in the right panel of Fig.~\ref{fig:Stiffness}(b)). The horizontal stiffness is presented for links oriented at different angles, and when $\eta=56^{\circ}$, which corresponds to one of the intersections of the $K_x$ and $K_y$ curves in the left panel of Fig.~\ref{fig:Stiffness}(b), we get a perfect circle meaning a uniform in-plane stiffness or an isotropic stiffness. The magnitude of the isotropic stiffness is still reasonably high compared to the highest $K_{xy}$ among all polar plots (the uniform stiffness is $6.9\times10^4$ N/m while the highest is $1.1\times10^5$ N/m). We denote the magnitude of the isotropic stiffness as $K_{xy}^*$ and the corresponding link orientation angle as $\eta^*$. We then explore the isotropic stiffness for basic assemblies with different Miura sector angles $\gamma$, and we can see that the specific link orientation angle for isotropic stiffness $\eta^*$ is always within the design space of rigid, flat foldable origami (see Fig.~\ref{fig:Stiffness}(c)). The magnitude of the isotropic stiffness $K_{xy}^*$ remains reasonably high regardless of the Miura sector angle, and thus it is always possible to design the link orientation angle if one wants to avoid an anisotropic behavior for in-plane loads. Figure~\ref{fig:Stiffness}(d) shows a physical prototype that demonstrates the load-bearing capacity with an isotropic design (the prototype is a $2\times3\times3$ block, and the link orientation angle is $56^{\circ}$, the value of $\eta^*$ when $\gamma=60^{\circ}$). 

Despite the ability to achieve isotropic stiffness, our multi-layered spaced origami would have an overall lower stiffness than other more continuous 3D Miura-ori assemblies such as stacked Miura-ori \citep{schenk2013geometry} and Miura-ori tube assemblies \citep{filipov2015origami}. The stiffness of our design is lower because the Miura-ori sheets are only connected by sparsely arranged thin-sheet links, while other 3D assemblies are densely packed with continuous internal material. Thus, the stiffness of the multi-layered spaced origami is strongly affected by the design parameters of these connector links. We use mechanical simulations to find that the isotropic stiffness $K_{xy}^*$ scales roughly with the link length as $d^{-1}$, the link width as $w^{\frac{4}{3}}$, and the link thickness as $t^{\frac{4}{3}}$ (see Sec. S3.2 of the Supplementary Material for details). Therefore, the in-plane stiffness can be increased by thicker materials and wider links, and it decreases linearly when the spacing between sheets increases. 

\section{Demonstration of potential applications}\label{sec:Applications}
As mentioned in Sec.~\ref{sec:GeometryDef}, the linkages between the sheets can be installed in any pattern without affecting the kinematics, and the Miura-ori can be arranged to create spaced sheets that geometrically approximate complex shapes. These properties bring great flexibility to engineering design. In this section, we give two examples to show how multi-layered spaced origami can be used to build functional devices for multi-physical engineering applications. These examples are based on existing metamaterial designs that have been verified by simulation and experimental testing \citep{popa2011experimental,NASA}. Here, we are providing an origami solution to fold and deploy these structures rather than proposing new metamaterial designs or principles. 

\subsection{Meter-scale acoustic cloak}
Evenly-spaced perforated plates can be used as an acoustic metamaterial to alter the effective properties of air and thus to achieve acoustic invisibility \citep{popa2011experimental}. Metamaterials of this type are known as \textit{acoustic cloaks}. Physical realizations of acoustic cloaks have all been at the centimeter scales \citep{fleury2015invisibility}. In this example, we use multi-layered spaced origami to design a foldable acoustic cloak at the meter scale. 

\begin{figure}[!htb] 
\centering
\makebox[0pt]{\includegraphics[scale=1]{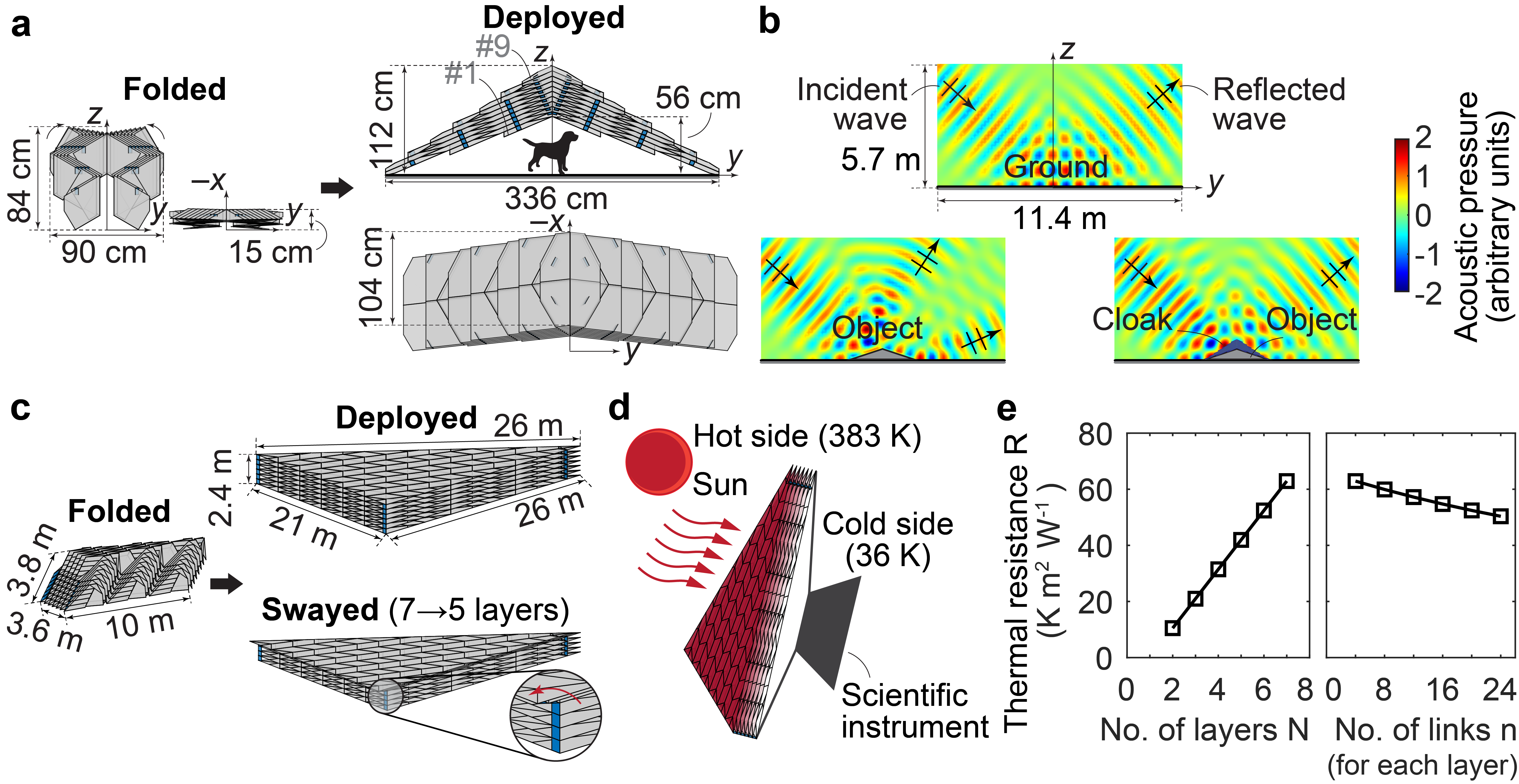}} 
\caption{Applications of multi-layered spaced origami. (a, b) Design and simulation of an acoustic cloak at the meter scale: (a) Dimensions of the folded and unfolded cloak; (b) Numerical testing of the cloak. (c, d, e) Design and simulation of an adjustable heat shield: (c) Dimensions of the folded, unfolded, and swayed shield; (d) An illustration of how the heat shield protects scientific instruments from high temperatures; (e) The thermal resistance $R$ changes with the number of layers $N$ and the number of links $n$ for each layer.}\label{fig:Application}
\end{figure}

The first step is to design the basic assembly. We choose $\eta=60^{\circ}$ as the link orientation angle to make the cloak flat foldable, rigid foldable, and a single-DOF mechanism. The values of all other parameters of the basic assembly are given in Sec. S4.1 of the Supplementary Material. The corresponding Miura-ori unit cell is then used to tessellate nine thin plates. The nine plates of different lengths are tilted to form a specific angle of $23.1^{\circ}$ with the ground and are evenly spaced at a distance $d=5$ cm (which is the same spacing as in our basic assembly). We drill circular holes of diameter $s=8$ mm at an interval $d$ in all nine plates, and assemble the cloak by connecting the nine perforated plates with 80 sparsely mounted links. Details on the origami patterns of the nine plates, the acoustic metamaterial unit cell design, and the entire deployment process of the cloaking structure can be found in Sec. S4.1 of the Supplementary Material. The folded and deployed states of the origami cloak are shown in Fig.~\ref{fig:Application}(a), and the packing ratio is 35. 

The acoustic cloak only has cloaking effects when the origami structure is fully deployed. Numerical simulation validates the performance of the acoustic cloak (see Fig.~\ref{fig:Application}(b)). The simulation is done in COMSOL Multiphysics\textsuperscript{®}. We perform a frequency domain analysis in the Pressure Acoustics module using a plane wave with a frequency of 300 Hz and a Gaussian-modulated amplitude as the incident wave. The origami panels are assumed to be made of acrylic plates that act as rigid walls and therefore do not interact elastically with acoustic waves. As such, the spaced sheets with holes modify the effective bulk modulus and density of air and make it suitable for acoustic cloaking \citep{popa2011experimental}. Compared to the perfect reflection in the pure ground case, severe scattering is caused by a triangular object placed on the ground, which becomes an acoustic signature and makes the object detectable. After we cover the object with the cloak, a near perfect reflection appears again with reflected waves appearing near identical to the pure ground case. We set the frequency to $f_0=$ 300 Hz in this example, but the cloak works well within a $f_0/3=$ 100 Hz bandwidth. More results and discussions on the broadband performance of this system can be found in Sec. S4.1 of the Supplementary Material. The designed cloak makes objects acoustically invisible at the meter-scale, and our origami design helps to achieve greater portability and modular fabrication of the structure. 

\subsection{Adjustable heat shield}
Multi-layer insulation is a common strategy to reduce heat transfer between two environments, especially in aerospace \citep{Incropera1996introduction}. In this example, we use multi-layered spaced origami to design a heat shield with adjustable thermal properties. We set the link orientation angle $\eta$ to $90^{\circ}$ so that in addition to the folding kinematics the basic assembly also has a sway degree of freedom (see Sec. S4.2 of the Supplementary Material for the values of all other design parameters). This sway motion will allow for varying the thermal properties of the system. As Fig.~\ref{fig:Application}(c) shows, each layer of our heat shield is tessellated with the Miura-ori unit cell and is a triangular `parasol' roughly the size of a tennis court. Seven layers in total are connected by the $90^{\circ}$-oriented links installed on three corners of each layer. 

The folded, unfolded, and swayed states of the heat shield are shown in Fig.~\ref{fig:Application}(c), and the packing ratio is 9.6 (see Sec. S4.2 of the Supplementary Material for the entire deployment process of the shielding structure). When the origami is fully deployed, the 7-layer structure limits heat transfer by attenuating thermal energy layer by layer. Assuming a vacuum environment in space and that the entire structure is made of highly reflective materials such as aluminum, we calculate the thermal resistance $R$ of the heat shield using \citep{Incropera1996introduction}
\begin{equation}\label{eqn:Thermal_R}
\begin{split}
R & = \frac{{{T_1} - {T_2}}}{{\underbrace {\frac{{\sigma (T_1^4 - T_2^4)}}{{\frac{2}{\varepsilon } - 1}}\frac{1}{{N - 1}}}_{q_{sheets}^{''}} + \underbrace {\frac{{nwtk}}{{(N - 1)dA}}({T_1} - {T_2})}_{q_{links}^{''}}}}\\
& = \dfrac{T_1-T_2}{\left ( \dfrac{a\left ( T_1^4-T_2^4 \right )}{\frac{2}{\varepsilon }-1}+\dfrac{nwtk\left ( T_1-T_2 \right )}{dA} \right )\dfrac{1}{N-1}},
\end{split}
\end{equation}
where $q_{sheets}^{''}$ and $q_{links}^{''}$ are the heat flux due to the radiation through the sheets and the conduction through the links; $T_1$, $T_2$ are the temperatures on the hot and cold sides; $N$ is the total number of layers while $n$ is the number of links for each layer; $A$ is the area of a single sheet; $d$, $w$, and $t$ are the length, width, and thickness of the links; $\sigma$ is the Stefan-Boltzmann constant, $\varepsilon$ is the emissivity of the sheets, and $k$ is the conductivity of the links. No heat convection exists, and the radiation and conduction are assumed to be uncoupled. 

According to Eq.~\ref{eqn:Thermal_R}, we can adjust the thermal resistance by changing the number of layers $N$, and this can be achieved by the sway mode of the designed origami. Fig.~\ref{fig:Application}(c) shows the 5-layer heat shield after the top two layers are collapsed by using a $90^{\circ}$ sway. We consider a real-life scenario where a scientific instrument in space needs a cool environment ($T_2=36\: \mathrm{K}$) to function properly while the sun ($T_1=383\: \mathrm{K}$) keeps transmitting thermal energy to the system (see Fig.~\ref{fig:Application}(d)) \citep{NASA}. For this scenario, we calculate the thermal resistance by substituting $A=249.75 \: \mathrm{m}^2$, $d=0.4\: \mathrm{m}$, $w=0.5\: \mathrm{m}$, $t=1\: \mathrm{mm}$, $\sigma=5.7\times10^{-8}\: \mathrm{W \:m^{-2}\:K^{-4}}$, $\varepsilon=0.05$ (for aluminum), and $k=237\: \mathrm{W \:m^{-1}\:K^{-1}}$ (for aluminum) into Eq.~\ref{eqn:Thermal_R}. Figure~\ref{fig:Application}(e) shows how the thermal resistance changes with the number of layers $N$ and with the number of links $n$ for each layer. We can adjust the thermal resistance over a wide range by collapsing layers through the sway motion (see the left panel in Fig.~\ref{fig:Application}(e)), and thus adaptively protect the delicate instrument from both extreme heat and cold. We can also enhance the stiffness of the shielding structure by installing more links without significantly affecting the thermal performance (see the right panel in Fig.~\ref{fig:Application}(e)). The links have a negligible effect on the thermal performance because the links are always sparsely installed, and the size of the link connections is negligible compared to the size of the sheets.

Despite the demonstrated performance of the acoustic cloak and heat shield, the results presented in this section have limitations because our work has made some assumptions regarding the simulations and we have not yet experimentally tested these systems. For example, in our work, we have ignored the initial deformation of the cloaking structure caused by its own weight, and we have also ignored the possible coupling of heat radiation and heat conduction during heat transfer through the shielding structure. Future work could investigate how these and other practical considerations affect the performance of metamaterials made of multi-layered spaced origami. 


\section{Conclusions}\label{sec:Conclusions}
This work presents a general framework for the design of multi-layered spaced origami, where separate Miura-ori sheets are connected with thin-sheet parallelogram linkages. We studied the kinematic and mechanical behavior of these multi-layered spaced origami systems by combining theoretical modeling, reduced-order FEM simulations, and physical experiments. The proposed modeling approach enables effective design and analysis of structures where multiple thin sheets are separately spaced and sparsely connected rather than bonded by continuous internal material. The multi-layered spaced origami has broad applications in multi-physical engineering scenarios such as architectural acoustics, space structures, and energy harvesting. Our main conclusions are summarized as follows: 
\begin{itemize}
    \item An analytical kinematic compatibility equation for rigid folding of multi-layered spaced origami is derived based on spatial trigonometry and rigid body transformation. This compatibility equation represents the quantitative relationship that couples the Miura-ori folding angle $\theta$ to the link rotation angle $\beta$ during the rigid folding motion of a multi-layered origami system. 
    \item The link orientation angle $\eta$ determines which of three unique solutions will satisfy the kinematic compatibility equation. These solutions represent three types of kinematic folding paths the origami system will follow including i) a flat foldable path where the system folds continuously from a deployed to a stowed state, ii) a self-locking path where the system folds continuously but will self-lock at a certain folding angle, and iii) a double-branch path where the system can have a sway motion between the spaced sheets yet can also fold continuously following a flat foldable path when $\eta=90^{\circ}$ or a self-locking path when $\eta=0^{\circ}$ or $180^{\circ}$. 
    \item The stiffness of the base pattern material does not have a significant influence on the folding behavior when the system folds along the flat-foldable or self-locking kinematic paths. In these scenarios, the elastic folding behavior matches well with the theoretically derived kinematics. On the other hand, in systems with a double-branch folding path (when $\eta=90^{\circ}$), the elastic behavior deviates from rigid folding. More flexible materials cause the system to jump from the sway branch to the flat-foldable branch when the system is subjected to a single-direction load instead of requiring loading in two different directions (which is typically needed to overcome bifurcation points). This behavior simplifies actuation of the origami system. 
    \item A larger spacing between sheets brings less interference during folding and thus enlarges the design space for rigid and flat foldable multi-layered spaced origami. However, a larger spacing also causes less compact packing. By strategically choosing the link orientation angle $\eta$, we can achieve the largest packing ratio when all other design parameters are fixed. 
    \item The folding mode of a multi-layered spaced origami system can be locked to convert the mechanism into a structure with shear and axial stiffness. For any Miura pattern geometry, we can find an optimal link orientation angle $\eta^*$ such that the structure has a uniform stiffness in all in-plane directions. These optimal solutions $\eta^*$ are all within the design space for rigid and flat foldability. The magnitude of the uniform stiffness $K_{xy}^*$ is reasonably high compared to the highest stiffness that can be achieved in all non-uniform-stiffness cases. 
    \item We designed and simulated an acoustic cloaking structure and a heat shielding structure at the meter-scale using the basic assemblies of multi-layered spaced origami. Both origami structures have a flat-foldable folding path, making them portable and modular. The heat shield is also adjustable regarding the thermal resistance thanks to the sway mode of the double-branch kinematic path when $\eta=90^{\circ}$. 
\end{itemize}

\section*{Acknowledgments}
The authors acknowledge support from the Automotive Research Center (ARC) in accordance with Cooperative Agreement W56HZV-19-2-0001 U.S. Army Ground Vehicle System Center (GVSC) in Warren, MI. The authors also acknowledge helpful discussions with Dr. Bogdan Popa, Dr. Yi Zhu and Mr. Haimiti Atila. The paper reflects the views and opinions of the authors, and not necessarily those of the funding entities. 





\end{spacing}

\setlength{\bibsep}{0.0pt} 




\end{document}